\documentclass[twocolumn,showpacs,showkeys,preprintnumbers,amsmath,amssymb]{revtex4}

\usepackage{graphicx}

\begin{document}


\title{Scaling laws in high-energy inverse Compton scattering. II. effect of bulk motions}
\author{Satoshi Nozawa}
 \email{snozawa@josai.ac.jp}
\affiliation{
Josai Junior College, 1-1 Keyakidai, Sakado-shi, Saitama, 350-0295,
Japan}

\author{Yasuharu Kohyama and Naoki Itoh}
\affiliation{
Department of Physics, Sophia University, 7-1 Kioi-cho, Chiyoda-ku,
Tokyo, 102-8554, Japan}

\date{\today}

\begin{abstract}
  We study the inverse Compton scattering of the CMB photons off high-energy nonthermal electrons.  We extend the formalism obtained by the previous paper to the case where the electrons have non-zero bulk motions with respect to the CMB frame.  Assuming the power-law electron distribution, we find the same scaling law for the probability distribution function $P_{1,K}(s)$ as $P_{1}(s)$ which corresponds to the zero bulk motions, where the peak height and peak position depend only on the power-index parameter.  We solved the rate equation analytically.  It is found that the spectral intensity function also has the same scaling law.  The effect of the bulk motions to the spectral intensity function is found to be small.  The present study will be applicable to the analysis of the X-ray and gamma-ray emission models from various astrophysical objects with non-zero bulk motions such as radio galaxies and astrophysical jets.
\end{abstract}



\maketitle

\section{Introduction}

  The inverse Compton scattering is one of the most fundamental reactions which have a variety of applications to astrophysics and cosmology.  They are, for example, the Sunyaev-Zeldovich (SZ) effects\cite{suny72} for clusters of galaxies (CG), cosmic-ray emission from radio galaxies\cite{blun06} and clusters of galaxies\cite{sara99}, and radio to gamma-ray emission from supernova remnants\cite{bari99, laze04}.  Therefore, theoretical studies on the inverse Compton scattering have been done quite extensively for the last forty years, starting from the works by Jones\cite{jone68}, and Blumenthal and Gould\cite{blum70} to the recent works, for example, by Fargion\cite{farg97}, Colafrancesco\cite{cola08, cola09}, and Petruk\cite{petr09}.

  In particular, remarkable progress has been made in theoretical studies for the SZ effects for CG.  Wright\cite{wrig79} and Rephaeli\cite{reph95} calculated the photon frequency redistribution function in the electron rest frame, which is called as the radiative transfer method.  On the other hand, Challinor and Lasenby\cite{chal98} and Itoh, Kohyama, and Nozawa\cite{itoh98} solved the relativistically covariant Boltzmann collisional equation for the photon distribution function, which is called the covariant formalism.  Although the two are very different approaches, the obtained results for the SZ effect agreed extremely well.  This has been a longstanding puzzle in the field of the relativistic study of the SZ effect for the last ten years.  Very recently, however, Nozawa and Kohyama\cite{noza09a} showed that the two formalisms were indeed mathematically equivalent in the approximation of the Thomson limit.  This explained the reason why the two different approaches produced the same results for the SZ effect even in the relativistic energies for electrons.  With the formalism, studies on various formal solutions and numerical solutions were presented\cite{noza09b}.

  Furthermore, Nozawa, Kohyama and Itoh\cite{noza10} (denoted paper I hereafter) applied the formalism obtained by Nozawa and Kohyama\cite{noza09a} to the inverse Compton scattering of the CMB photons off high-energy nonthermal electrons.  This extension is particularly interesting for the analysis of X-ray and gamma-ray emissions, for example, from radio galaxies\cite{blun06} and supernova remnants\cite{bari99, laze04}, where the inverse Compton scattering of the CMB photons off nonthermal high-energy electrons plays an essential role.  In paper I, a universal scaling law was shown for the redistribution function $P_{1}(s)$ and the spectral intensity function $I(x)$ under a specific condition for the electron distribution which is typically realized.  It was shown that the spectral intensity function for different energy scales (, for example, keV, MeV and GeV) were described by one equation with a scaling variable $X=x/4\gamma_{min}^{2}$, where $x$ is the photon energy in units of the thermal energy of the CMB and $\gamma_{min}$ is the minimum value of the Lorentz factor of the electron power-law distribution.

  In the present paper, we extend the formalism obtained by paper I to the electrons of the astrophysical objects with bulk motions.  As for the CG, the effect of the bulk motions was originally obtained by Sunyaev and Zeldovich\cite{suny80}, which is known as the kinematical SZ effect.  The relativistic corrections to the kinematical SZ effect was presented by Nozawa, Itoh and Kohyama\cite{noza98}.  We will explore the effect of the bulk motions of the astrophysical objects (, for example, the peculiar velocity of the CG) for the high-energy inverse Compton scattering.  This extension will be particularly interesting for the analysis of X-ray and gamma-ray emissions, for example, from radio galaxies with non-zero peculiar velocities and various astrophysical jets.

  On the other hand, it is well known that the Solar System (, i.e., the observer) has a bulk motion with respect to the CMB frame.  Assuming that the CMB dipole is fully motion-induced, we deduce that the Solar System is moving with a velocity $\beta_{S} \equiv v_{S}/c$ = 1.241$\times10^{-3}$ towards the direction $(\ell, b)=(264.14^{\circ} \pm 0.15^{\circ}$, $48.26^{\circ} \pm 0.15^{\circ}$)\cite{smoo77, fixs96, fixs02}.  Chluba et al.\cite{chlu05} calculated the corrections to the SZ effect for the CG arising from the bulk motion of the Solar System.  Nozawa, Itoh and Kohyama\cite{noza05} calculated the effect in more general way with the Lorentz covariant formalism.  In the present paper, we apply the method developed by Nozawa, Itoh and Kohyama\cite{noza05} to the high-energy inverse Compton scattering and calculate the effect of the bulk motion of the observer.  Thus, the effect of the bulk motions for both the astrophysical object and observer will be derived in the present paper.

  Before closing the present section, it should be emphasized the following:  In the present approach, we push analytic techniques as much as possible in order to obtain analytic solutions.  In contrast to the direct numerical calculation, the present approach will have an advantage that one may reveal essential physics properties behind the numerical results.  In the present paper, under a specific condition for the electron distribution which is typically realized, we will show that a universal scaling law is established for the spectral intensity function even if one includes the effect of the bulk motions for both the astrophysical object and observer.

  The present paper is organized as follows:  In Sec.~II, we derive the analytic expressions for the redistribution functions $P_{K}(s,\gamma)$ and $P_{1,K}(s)$.  Assuming the power-law electron distribution, we show that $P_{1,K}(s)$ has the same scaling law as $P_{1}(s)$, where the peak height and peak position depend only on the power-index parameter.  We calculate the rate equation and obtain the analytic expression for the spectral intensity function $dI(X)/d\tau$.  We show that $dI(X)/d\tau$ also has the scaling law, where the peak height and peak position depend only on the power-index parameter.  In Sec.~III, we derive the the analytic expression for the spectral intensity function $dI(X)/d\tau$ which includes the effect of the bulk motion of the observer.  Finally, concluding remarks are given in Sec.~IV.

\section{High-energy Inverse Compton Scattering off Electrons with Bulk Motion}

\subsection{Rate equations in the Thomson approximation}

  In Nozawa and Kohyama\cite{noza09a}, it was shown that the covariant formalism\cite{itoh98} and radiative transfer method\cite{wrig79} were mathematically equivalent in the following (Thomson) approximation:
\begin{eqnarray}
&&\hspace{-10mm}
\gamma \frac{\omega}{m} \ll 1  \, ,
\label{eq2a-1}   \\
&&\hspace{-10mm}
\gamma = \frac{1}{\sqrt{1-\beta^{2}}}  \, ,
\label{eq2a-2}
\end{eqnarray}
where $\omega$ is the photon energy, $\gamma$ is the Lorentz factor, and $\beta$ and $m$ are the velocity and rest mass of the electron, respectively.  Throughout this paper, we use the natural unit $\hbar = c = 1$, unless otherwise stated explicitly.  For the CMB photons, Eq.~(\ref{eq2a-1}) is fully valid from nonrelativistic electrons to extreme-relativistic electrons of the order of TeV region.  In paper I\cite{noza10}, the high-energy inverse Compton scattering of the CMB photons has been studied under the assumption of Eq.~(\ref{eq2a-1}).

  In the present paper, we extend the formalism obtained in paper I to the electron distribution with a bulk motion.  Let us suppose that the astrophysical object (, for example, we consider the CG in the present paper) is moving with a bulk velocity $\vec{\beta}_{C}$ (=$\vec{v}_{C}/c$) with respect to the CMB frame.  As a reference system, we choose the system that is fixed to the CMB in the present section.  We discuss the effect of the bulk motion of the observer (the Solar System) in Sec.~III.  The $z$ axis is fixed to a line connecting the observer and the center of mass of the CG.  (We assume that the observer is fixed to the CMB frame.)  In the present paper we choose the positive direction of the $z$ axis as the direction of the propagation of a photon from the observer to the CG.

  The rate equations for the photon distribution function $n(x)$ and spectral intensity function $I(x)$ were derived in Nozawa and Kohyama\cite{noza09a} under the assumption of Eq.~(\ref{eq2a-1}).  Here, $x=\omega/k_{B}T_{CMB}$ is the photon energy in units of the thermal energy of the CMB, and $s$ is the frequency shift defined by $e^{s}=x^{\prime}/x$.  We recall the results here to make the present paper more self-contained.  They are given as follows\cite{noza09a, noza09b}:
\begin{eqnarray}
&&\hspace{-10mm}
\frac{\partial n(x)}{\partial \tau}
 = \int_{-\infty}^{\infty}ds P_1(s,\beta_{C,z})
\left[n(e^sx)- n(x)\right] \, ,
\label{eq2a-3}   \\
&&\hspace{-10mm}
\frac{\partial I(x)}{\partial \tau}
 = \int_{-\infty}^{\infty}ds
{P}_1(s,\beta_{C,z}) \left[e^{-3s}I(e^{s}x)- I(x)\right]  \, ,
\label{eq2a-4}  \\
&&\hspace{+12mm}
d\tau  =  n_e\sigma_T dt \, , 
\label{eq2a-5}  \\
&&\hspace{0mm}
P_1(s,\beta_{c,z}) = P_{1}(s) + \beta_{C,z} P_{1, K}(s)  \, ,
\label{eq2a-6}
\end{eqnarray}
where $I(x)=I_{0}x^{3}n(x)$, $I_{0}=(k_{B}T_{CMB})^{3}/2\pi^{2}$, $n_{e}$ is the electron number density, $\sigma_{T}$ is the Thomson scattering cross section, and $\beta_{C,z}$ is the bulk velocity of the CG parallel to the observer.  It should be noted that $O(\beta_{C,z}^{2})$ and higher-order contributions were neglected in deriving Eq.~(\ref{eq2a-6}), because $\beta_{C,z} \ll 1$ is satisfied for most of the CG.

  In Eq.~(\ref{eq2a-6}), $P_{1}(s)$ was calculated in paper I, and $P_{1,K}(s)$ is the term which appears in the case of non-zero bulk motions.  They are defined as follows:
\begin{eqnarray}
&&\hspace{-11mm}
P_1(s) = \int_{\beta_{min}}^{1}d\beta\beta^2\gamma^5 p_e(E)P(s,\beta)  \, ,
\label{eq2a-7}  \\
&&\hspace{-11mm}
 P_{1,K}(s) = \int_{\beta_{min}}^{1}d\beta\beta^2\gamma^5 p_e(E)P_{K}(s,\beta)  \, ,
\label{eq2a-8}  \\
&&\hspace{-11mm}
P(s,\beta) = \frac{e^{s}}{2\beta\gamma^4}
\int_{\mu_1(s)}^{\mu_2(s)}d\mu_0 \frac{1}{(1-\beta\mu_0)^2} f\left(\mu_0, \mu_0^{\prime} \right)   \, ,
\label{eq2a-9}  \\
&&\hspace{-12mm}
P_{K}(s,\beta)
= \frac{e^{s}}{2\beta\gamma^4} \delta(\beta)
\int_{\mu_1(s)}^{\mu_2(s)}d\mu_0 \frac{\beta \mu_{0}-\beta^2}{(1-\beta\mu_0)^3}
f\left(\mu_0, \mu_{0}^{\prime} \right)  ,
\label{eq2a-10}  \\
&&\hspace{-11mm}
f(\mu_0,\mu_0^{\prime}) = \frac{3}{8}\left[
  1 + \mu_0^2\mu_0^{\prime 2}+\frac{1}{2}(1-\mu_0^2)(1-\mu_0^{\prime 2})
\right]  \, .
\label{eq2a-11}
\end{eqnarray}
In Eq.~(\ref{eq2a-10}), $\delta(\beta)$ is a factor related to the electron distribution function, which is, in general, a function of $\beta$.  The explicit forms are given by Nozawa and Kohyama\cite{noza09a} for three different electron distribution functions.  We will define the explicit form later in this section.  The electron distribution function of a momentum $p$ is normalized by $\int_{0}^{\infty} dp p^{2} p_{e}(E)/m^{3}=1$.  Variables appearing in Eqs.~(\ref{eq2a-7}) -- (\ref{eq2a-11}) are summarized as follows:
\begin{eqnarray}
&&\hspace{-10mm}
\beta_{min} = (1-e^{-|s|})/(1+e^{-|s|})  \, ,
\label{eq2a-12} \\
&&\hspace{-10mm}
\mu_{0}^{\prime} = [1-e^s(1-\beta\mu_0)]/\beta  \, ,
\label{eq2a-13}  \\
&&\hspace{-10mm}
\mu_1(s) = \left\{
\begin{array}{ll}
-1 &\quad  {\rm for} \, \, \, s \leq 0 \\
{[1-e^{-s}(1+\beta)]/\beta} &\quad {\rm for} \, \, \, s > 0
\end{array}
\right.  \, ,
\label{eq2a-14} \\
&&\hspace{-10mm}
\mu_2(s) = \left\{
\begin{array}{ll}
{[1-e^{-s}(1-\beta)]/\beta} &\quad {\rm for} \, \, \, s < 0 \\
1 &\quad  {\rm for} \, \, \, s \geq 0 
\end{array}
\right. \, .
\label{eq2a-15}
\end{eqnarray}

  The total probabilities for $P(s,\beta)$ and $P_{K}(s,\beta)$ are given by
\begin{eqnarray}
&&\hspace{-15mm}
\int_{-\lambda_{\beta}}^{+\lambda_{\beta}}ds P(s, \beta) = 1
\label{eq2a-16}  \, , \\
&&\hspace{-15mm}
\int_{-\lambda_{\beta}}^{+\lambda_{\beta}}ds P_{K}(s, \beta) = \frac{1}{3} \delta (\beta) \beta^{2}
\label{eq2a-17}  \, ,
\end{eqnarray}
where
\begin{eqnarray}
&&\hspace{-10mm}
\lambda_{\beta} = \ln \left(\frac{1+\beta}{1-\beta} \right)
\label{eq2a-18}  \, .
\end{eqnarray}
It should be noted that the following useful relations:
\begin{eqnarray}
&&\hspace{-10mm}
P(s, \beta)e^{-3s} = P(-s,\beta)
\label{eq2a-19}  \, , \\
&&\hspace{-7mm}
P_{1}(s)e^{-3s} = P_{1}(-s)
\label{eq2a-20}
\end{eqnarray}
are valid for $P(s,\beta)$ and $P_{1}(s)$.

\subsection{$P_{K}(s, \beta)$ for extreme-relativistic electrons}

 In this section, we derive the analytic expression of the frequency redistribution function $P_{K}(s, \beta)$ for extreme-relativistic electrons.  In order to proceed the calculation, we rewrite Eq.~(\ref{eq2a-10}) as follows:
\begin{eqnarray}
&&\hspace{-10mm}
P_{K}(s,\beta) = \delta(\beta) \left[ \tilde{P}_{K}(s, \beta) - P(s, \beta) \right]  \, ,
\label{eq2b-1}  \\
&&\hspace{-10mm}
\tilde{P}_{K}(s,\beta)
= \frac{e^{s}}{2\beta\gamma^6}
\int_{\mu_1(s)}^{\mu_2(s)}d\mu_0
\frac{1}{(1-\beta\mu_0)^3}
f\left(\mu_0, \mu_{0}^{\prime} \right)   \, ,
\label{eq2b-2}
\end{eqnarray}
where $P(s,\beta)$ in Eq.~(\ref{eq2b-1}) is defined by Eq.~(\ref{eq2a-9}) and the explicit forms were derived in paper I. (Readers may be referred to paper I for the explicit forms.)  Therefore, the main concern in this section is to derive the explicit forms for $\tilde{P}_{K}(s,\beta)$.  Note that the identity relation
\begin{eqnarray}
&&\hspace{-10mm}
\frac{\beta \mu_{0} - \beta^{2}}{\left(1 - \beta \mu_{0} \right)^{3}} = \frac{1}{\gamma^{2} \left(1 - \beta \mu_{0} \right)^{3}} -  \frac{1}{\left(1 - \beta \mu_{0} \right)^{2}}
\label{eq2b-3}
\end{eqnarray}
was used in deriving Eq.~(\ref{eq2b-1}).  It is also important to mention that Eq.~(\ref{eq2b-2}) satisfies the following relation:
\begin{eqnarray}
&&\hspace{-10mm}
\tilde{P}_{K}(s, \beta)e^{-4s} = \tilde{P}_{K}(-s,\beta)
\label{eq2b-4}  \, .
\end{eqnarray}

  In Eq.~(\ref{eq2b-2}), the integral of $\mu_{0}$ can be done analytically.  One obtains as follows: for $s<0$,
\begin{eqnarray}
&&\hspace{-10mm}
\tilde{P}_{K}(s,\beta) = \frac{3}{32\beta^2\gamma^4}
\left[
A_1(\beta)e^s + A_2(\beta)e^{2s} + A_3(\beta)e^{3s}
\right.
\nonumber   \\
&&\hspace{-5mm}
\left.
+ \left( \lambda_{\beta}+s \right) \left( B_1(\beta)e^s + B_2(\beta)e^{2s}
+ B_1(\beta)e^{3s} \right)
\right]  \, ,
\label{eq2b-5}
\end{eqnarray}
and for $s\ge0$,
\begin{eqnarray}
&&\hspace{-10mm}
\tilde{P}_{K}(s,\beta) = \frac{3}{32\beta^2\gamma^4}
\left[
A_3(\beta)e^s + A_2(\beta)e^{2s} + A_1(\beta)e^{3s}
\right.
\nonumber   \\
&&\hspace{-5mm}
\left. + \left( \lambda_{\beta}-s \right) \left( B_1(\beta)e^s + B_2(\beta)e^{2s} + B_1(\beta)e^{3s} \right)
\right]  \, ,
\label{eq2b-6}
\end{eqnarray}
where the coefficients are
\begin{eqnarray}
&&\hspace{-14mm}
A_1(\beta) = \frac{1}{\beta^4\gamma^2(1+\beta)^{2}} 
\left(9 + 12\beta - 4\beta^{2} - 8 \beta^{3} - 3 \beta^{4} \right)
\label{eq2b-7} \, , \\
&&\hspace{-14mm}
A_2(\beta) = \frac{8}{\beta^{3}} \left(-3 + 2 \beta^{2} \right)
\label{eq2b-8} \, , \\
&&\hspace{-14mm}
A_3(\beta) =  \frac{1}{\beta^4\gamma^2(1-\beta)^{2}} 
\left(-9 + 12\beta + 4\beta^{2} - 8 \beta^{3} + 3 \beta^{4} \right)
\label{eq2b-9} \, ,
\end{eqnarray}
\begin{eqnarray}
&&\hspace{-14mm}
B_1(\beta) = \frac{(3-\beta^2)}{\beta^4\gamma^2}
\label{eq2b-10} \, ,  \\
&&\hspace{-14mm}
B_2(\beta) = \frac{12}{\beta^{4} \gamma^{2}}
\label{eq2b-11} \, .
\end{eqnarray}
It is clear that Eqs.~(\ref{eq2b-5}) and (\ref{eq2b-6}) satisfy the relation of Eq.~(\ref{eq2b-4}).

  Now let us consider the case for electrons of extreme-relativistic energies $E$ $(=\gamma mc^{2}) \gg mc^{2}$.  Thus, $\gamma \gg 1$ and $\beta \approx 1$ are assumed.  Under this approximation, one can rewrite Eqs.~(\ref{eq2b-5}) and (\ref{eq2b-6}), and one finally obtains as follows: for $s<0$,
\begin{eqnarray}
&&\hspace{-15mm}
\tilde{P}_{K}(s,\gamma) = \frac{3}{32\gamma^4}
\Bigl[ \frac{3e^{s}}{2\gamma^2} - 8e^{2s} + 8e^{3s} \gamma^{2}
\nonumber  \\
&&\hspace{+15mm}
+ \left(\lambda_{\gamma}+s\right)\frac{2e^s}{\gamma^{2}} \Bigr]
\label{eq2b-12} \, ,
\end{eqnarray}
and for $s\ge0$,
\begin{eqnarray}
&&\hspace{-15mm}
\tilde{P}_{K}(s,\gamma) = \frac{3}{32\gamma^4}
\Bigl[ 8e^{s}\gamma^{2} - 8e^{2s} + \frac{3e^{3s}}{2\gamma^2}
\nonumber  \\
&&\hspace{+15mm}
+ \left(\lambda_{\gamma}-s\right)\frac{2e^{3s}}{\gamma^{2}} \Bigr]
\label{eq2b-13} \, ,
\end{eqnarray}
where 
\begin{eqnarray}
&&\hspace{-10mm}
\lambda_{\gamma} = 2 {\rm ln}(2\gamma)
\label{eq2b-14} \, ,
\end{eqnarray}
and the expression $\tilde{P}_{K}(s,\gamma)$ was used instead of $\tilde{P}_{K}(s,\beta)$.  It should be noted that only leading-order terms were kept in deriving Eq.~(\ref{eq2b-13}), and the relation of Eq.~(\ref{eq2b-4}) was applied in deriving Eq.~(\ref{eq2b-12}).  Thus, one has the total probabilities
\begin{eqnarray}
&&\hspace{-10mm}
\int_{-\lambda_{\gamma}}^{+\lambda_{\gamma}}ds \tilde{P}_{K}(s, \gamma) = \frac{4}{3} + O\left(\frac{1}{\gamma^{2}}\right)
\label{eq2b-15}  \, , \\
&&\hspace{-10mm}
\int_{-\lambda_{\gamma}}^{+\lambda_{\gamma}}ds P_{K}(s, \gamma) = \frac{1}{3}\delta(\gamma) + O\left(\frac{1}{\gamma^{2}}\right)
\label{eq2b-16}  \, ,
\end{eqnarray}
where the expression $\delta(\gamma)$ was used instead of $\delta(\beta)$.

\subsection{Scaling law of $P_{1,K}(s)$ for nonthermal electrons}

  In order to proceed calculation for practical applications, let us specify the electron distribution function.  High-energy electrons in the supernova remnants and active galactic nuclei, for example, are most likely nonthermal.  It is standard to describe the nonthermal distribution in terms of the power-law distribution function of three parameters:
\begin{eqnarray}
&&\hspace{-10mm}
p_{e}(\gamma) = \left\{
\begin{array}{ll}
 N_{\gamma} \, \gamma^{-\sigma} \, , &  \, \, \, \gamma_{min} \leq \gamma \leq \gamma_{max} \\
   0  \, , &  \, \, \, {\rm elsewhere}
\end{array}
\right.  \, ,
\label{eq2c-1}
\end{eqnarray}
where $\gamma$ is the Lorentz factor and $N_{\gamma}$ is the normalization constant.  In Eq.~(\ref{eq2c-1}), $\sigma$ is the power-index parameter, $\gamma_{min}$ and $\gamma_{max}$ are parameters of minimum and maximum values for $\gamma$, respectively.  It is known by Nozawa and Kohyama\cite{noza09a} that the factor $\delta(\beta)$ is a constant which depends only on the power-index in the case of the $\gamma$-power distribution of Eq.~(\ref{eq2c-1}).  Therefore, we use the expression $\delta$ instead of $\delta(\beta)$ hereafter.

Then, using the definition of Eq.~(\ref{eq2b-1}), Eq.~(\ref{eq2a-8}) can be reexpressed as follows:
\begin{eqnarray}
&&\hspace{-10mm}
P_{1,K}(s) = \delta \left[ \tilde{P}_{1,K}(s) - P_{1}(s) \right]  \, ,
\label{eq2c-2}
\end{eqnarray}
where $P_{1}(s)$ is defined by Eq.~(\ref{eq2a-7}) and the explicit forms were derived in paper I.  In Eq.~(\ref{eq2c-2}), the function $\tilde{P}_{1,K}(s)$ is defined as follows: for $s<0$,
\begin{eqnarray}
&&\hspace{-10mm}
\tilde{P}_{1,K}(s) = \int_{\max(\gamma_{min},e^{-s/2}/2)}^{\gamma_{max}}d\gamma p_e(\gamma) \tilde{P}_{K}(s,\gamma)  \, ,
\label{eq2c-3}
\end{eqnarray}
where $\tilde{P}_{1,K}(s,\gamma)$ is given by Eq.~(\ref{eq2b-12}), and for $s\ge0$,
\begin{eqnarray}
&&\hspace{-10mm}
\tilde{P}_{1,K}(s) = \int_{\max(\gamma_{min},e^{s/2}/2)}^{\gamma_{max}}d\gamma p_e(\gamma)\tilde{P}_{K}(s,\gamma)  \, ,
\label{eq2c-4}
\end{eqnarray}
where $\tilde{P}_{K}(s,\gamma)$ is given by Eq.~(\ref{eq2b-13}).  In deriving Eqs.~(\ref{eq2c-3}) and (\ref{eq2c-4}), $\beta \approx 1$ was assumed, and the phase space factor $\gamma^{2}$ was absorbed, for simplicity, by the power-index $\sigma$ in $p_e(\gamma)$.  Since the explicit form for the electron distribution function has been fixed, the parameter $\delta$ is now determined as follows\cite{noza09a}:
\begin{eqnarray}
&&\hspace{-10mm}
\delta = \sigma + 2
\label{eq2c-5}  \, .
\end{eqnarray}

  In the case of the power-law distribution of Eq.~(\ref{eq2c-1}), Eqs~(\ref{eq2c-3}) and (\ref{eq2c-4}) can be integrated analytically.  The explicit forms are given as follows: for $-2\ln 2\gamma_{max}<s <-2\ln 2\gamma_{min} $,
\begin{eqnarray}
&&\hspace{-5mm}
\tilde{P}_{1,K}(s) = \frac{3}{32}N_{\gamma} \left[
\frac{e^{s}}{\sigma+5} \left\{ \frac{3 \sigma+23}{2(\sigma+5)} 2^{\sigma+5}e^{(\sigma+5)s/2}  \right. \right.
\nonumber  \\
&&\hspace{4mm}
\left. - \frac{1}{\gamma_{max}^{\sigma+5}} \left( \frac{3 \sigma+23}{2(\sigma+5)} +2s + 4\ln 2\gamma_{max} \right) \right\}
\nonumber \\
&&\hspace{4mm}
- \frac{8e^{2s}}{\sigma+3} \left(2^{\sigma+3}e^{(\sigma+3)s/2} - \frac{1}{\gamma_{max}^{\sigma+3}} \right)
\nonumber \\
&&\hspace{4mm}
\left.
+ \frac{8e^{3s}}{\sigma+1} \left( 2^{\sigma+1}e^{(\sigma+1)s/2}
- \frac{1}{\gamma_{max}^{\sigma+1}}\right) 
\right]
\label{eq2c-6}  \, ,
\end{eqnarray}
for $-2\ln 2\gamma_{min}<s<0$,
\begin{eqnarray}
&&\hspace{-5mm}
\tilde{P}_{1,K}(s) = \frac{3}{32}N_{\gamma} \left[
\frac{e^{s}}{\sigma+5} \left\{
\left(\frac{3 \sigma+23}{2(\sigma+5)} +2s \right)  \right. \right.
\nonumber \\
&&\hspace{4mm}
\left. \times \left(\frac{1}{\gamma_{min}^{\sigma+5}}-\frac{1}{\gamma_{max}^{\sigma+5}}\right) + \frac{4}{\gamma_{min}^{\sigma+5}} \ln \left( \frac{\gamma_{min}}{\gamma_{max}} \right) \right\}
\nonumber \\
&&\hspace{4mm}
- \frac{8e^{2s}}{\sigma+3} \left(\frac{1}{\gamma_{min}^{\sigma+3}} - \frac{1}{\gamma_{max}^{\sigma+3}} \right)
\nonumber \\
&&\hspace{4mm}
\left.
+ \frac{8e^{3s}}{\sigma+1} \left( \frac{1}{\gamma_{min}^{\sigma+1}}
- \frac{1}{\gamma_{max}^{\sigma+1}}\right) 
\right]
\label{eq2c-7}  \, ,
\end{eqnarray}
for $0 < s < 2\ln 2\gamma_{min}$,
\begin{eqnarray}
&&\hspace{-5mm}
\tilde{P}_{1,K}(s) = \frac{3}{32}N_{\gamma} \left[
\frac{e^{3s}}{\sigma+5} \left\{
\left(\frac{3 \sigma+23}{2(\sigma+5)} -2s \right)  \right. \right.
\nonumber \\
&&\hspace{4mm}
\left. \times \left(\frac{1}{\gamma_{min}^{\sigma+5}}-\frac{1}{\gamma_{max}^{\sigma+5}}\right) + \frac{4}{\gamma_{min}^{\sigma+5}} \ln \left( \frac{\gamma_{min}}{\gamma_{max}} \right) \right\}
\nonumber \\
&&\hspace{4mm}
- \frac{8e^{2s}}{\sigma+3} \left(\frac{1}{\gamma_{min}^{\sigma+3}} - \frac{1}{\gamma_{max}^{\sigma+3}} \right)
\nonumber \\
&&\hspace{4mm}
\left.
+ \frac{8e^{s}}{\sigma+1} \left( \frac{1}{\gamma_{min}^{\sigma+1}}
- \frac{1}{\gamma_{max}^{\sigma+1}}\right) 
\right]
\label{eq2c-8}  \, ,
\end{eqnarray}
and for $2\ln 2\gamma_{min}<s <2\ln 2\gamma_{max} $,
\begin{eqnarray}
&&\hspace{-5mm}
\tilde{P}_{1,K}(s) = \frac{3}{32}N_{\gamma} \left[
\frac{e^{3s}}{\sigma+5} \left\{ \frac{3 \sigma+23}{2(\sigma+5)} 2^{\sigma+5}e^{-(\sigma+5)s/2}  \right. \right.
\nonumber  \\
&&\hspace{4mm}
\left. - \frac{1}{\gamma_{max}^{\sigma+5}} \left( \frac{3 \sigma+23}{2(\sigma+5)} -2s + 4\ln 2\gamma_{max} \right) \right\}
\nonumber \\
&&\hspace{4mm}
- \frac{8e^{2s}}{\sigma+3} \left(2^{\sigma+3}e^{-(\sigma+3)s/2} - \frac{1}{\gamma_{max}^{\sigma+3}} \right)
\nonumber \\
&&\hspace{4mm}
\left.
+ \frac{8e^{s}}{\sigma+1} \left( 2^{\sigma+1}e^{-(\sigma+1)s/2}
- \frac{1}{\gamma_{max}^{\sigma+1}}\right) 
\right]
\label{eq2c-9}  \, .
\end{eqnarray}
It should be noted that the normalization constant is given by
\begin{eqnarray}
&&\hspace{-10mm}
N_{\gamma} = (\sigma-1)\gamma_{min}^{\sigma-1}
\label{eq2c-10}
\end{eqnarray}
for the case $\gamma_{\max} \to \infty$.

  Let us now introduce new functions $\tilde{P}_{K,C}(s,R)$ and $\tilde{P}^{\prime}_{K,IC}(s,R)$ in order to express Eqs.~(\ref{eq2c-6})--(\ref{eq2c-9}) in unified forms, where $R=\gamma_{min}/\gamma_{max}$.  Here, $C$ and $IC$ denote the Compton scattering and inverse Compton scattering, respectively.  First, we define $\tilde{P}_{K,C}(s,R)$ as follows: for $-2\ln 2\gamma_{min} < s < 0$,
\begin{eqnarray}
&&\hspace{-5mm}
\tilde{P}_{K,C}(s,R) = \frac{3(\sigma-1)}{1-R^{\sigma-1}}
\left[
\frac{e^{3s}}{\sigma+5} \left\{ \left(\frac{3 \sigma+23}{\sigma+5} -4s \right)  \right. \right.
\nonumber  \\
&&\hspace{4mm}
  \times \left(1 - R^{\sigma+5} \right) + 8R^{\sigma+5} \ln R \biggr\}
\nonumber \\
&&\hspace{4mm}
\left. - \frac{4e^{2s}}{\sigma+3} \left(1 - R^{\sigma+3} \right)
+ \frac{e^{s}}{\sigma+1} \left(1 - R^{\sigma+1} \right) 
\right]
\label{eq2c-11} \, ,
\end{eqnarray}
and for $0<s <2\ln (\gamma_{max}/\gamma_{min}) $,
\begin{eqnarray}
&&\hspace{-5mm}
\tilde{P}_{K,C}(s,R) = \frac{3(\sigma-1)}{1-R^{\sigma-1}}
\left[
\frac{e^{3s}}{\sigma+5} \left\{ \frac{3 \sigma+23}{\sigma+5} e^{-(\sigma+5)s/2}  \right. \right.
\nonumber  \\
&&\hspace{4mm}
\left. - R^{\sigma+5} \left( \frac{3 \sigma+23}{\sigma+5} - 4s - 8\ln R \right) \right\}
\nonumber \\
&&\hspace{4mm}
- \frac{4e^{2s}}{\sigma+3} \left(e^{-(\sigma+3)s/2} - R^{\sigma+3} \right)
\nonumber \\
&&\hspace{4mm}
\left.
+ \frac{e^{s}}{\sigma+1} \left( e^{-(\sigma+1)s/2} - R^{\sigma+1}\right) 
\right]
\label{eq2c-12} \, .
\end{eqnarray}
Similarly, $\tilde{P}^{\prime}_{K,IC}(s,R)$ is for $-2\ln (\gamma_{max}/\gamma_{min})< s < 0 $,
\begin{eqnarray}
&&\hspace{-5mm}
\tilde{P}^{\prime}_{K,IC}(s,R) = \frac{3(\sigma-1)}{1-R^{\sigma-1}}
\left[
\frac{e^{s}}{\sigma+5} \left\{ \frac{3 \sigma+23}{\sigma+5} e^{(\sigma+5)s/2}  \right. \right.
\nonumber  \\
&&\hspace{4mm}
\left. - R^{\sigma+5} \left( \frac{3 \sigma+23}{\sigma+5} + 4s - 8\ln R \right) \right\}
\nonumber \\
&&\hspace{4mm}
- \frac{4e^{2s}}{\sigma+3} \left(e^{(\sigma+3)s/2} - R^{\sigma+3} \right)
\nonumber \\
&&\hspace{4mm}
\left.
+ \frac{e^{3s}}{\sigma+1} \left( e^{(\sigma+1)s/2} - R^{\sigma+1}\right) 
\right]
\label{eq2c-13} \, ,
\end{eqnarray}
and for $0<s <2\ln 2\gamma_{min} $,
\begin{eqnarray}
&&\hspace{-10mm}
\tilde{P}^{\prime}_{K,IC}(s,R) = \frac{3(\sigma-1)}{1-R^{\sigma-1}}
\left[
\frac{e^{s}}{\sigma+5} \left\{ \left(\frac{3 \sigma+23}{\sigma+5} +4s \right)  \right. \right.
\nonumber  \\
&&\hspace{4mm}
  \times \left(1 - R^{\sigma+5} \right) + 8R^{\sigma+5} \ln R \biggr\}
\nonumber \\
&&\hspace{4mm}
\left. - \frac{4e^{2s}}{\sigma+3} \left(1 - R^{\sigma+3} \right)
+ \frac{e^{3s}}{\sigma+1} \left(1 - R^{\sigma+1} \right) 
\right]
\label{eq2c-14} \, .
\end{eqnarray}
It is straightforward to show that
\begin{eqnarray}
&&\hspace{-10mm}
\tilde{P}_{K,C}(s,R) e^{-4s} = \tilde{P}^{\prime}_{K,IC}(-s,R)
\label{eq2c-15}
\end{eqnarray}
is satisfied by Eqs.~(\ref{eq2c-11})--(\ref{eq2c-14}).

  Comparing Eqs.~(\ref{eq2c-6})--(\ref{eq2c-9}) with Eqs.~(\ref{eq2c-11})--(\ref{eq2c-14}), the probability distribution function $\tilde{P}_{1,K}(s)$ is described as follows:
\begin{eqnarray}
&&\hspace{-13mm}
\tilde{P}_{1,K}(s) = 
\left\{
\begin{array}{ll}
\tilde{P}_{K,IC}(s+2\ln 2\gamma_{min},R) &   {\rm for \,} s < 0 \\
  \\
\tilde{P}_{K,C}(s-2\ln 2\gamma_{min},R)  & {\rm for \,} s \ge 0
\end{array}
\right.
\label{eq2c-16} \, ,
\end{eqnarray}
where
\begin{eqnarray}
&&\hspace{-10mm}
\tilde{P}_{K,IC}(s,R) \equiv \frac{1}{256\gamma_{min}^8} \tilde{P}^{\prime}_{K,IC}(s,R)
\label{eq2c-17}  \, .
\end{eqnarray}
Combining the result of Eq.~(\ref{eq2c-16}) for $\tilde{P}_{1,K}(s)$ with the previous result for $P_{1}(s)$ in paper I, one finally obtains the total contribution $P_{1,K}(s)$ defined by Eq.~(\ref{eq2c-2}).

  Let us now consider the case $R \equiv \gamma_{min}/\gamma_{max} \ll 1$.  We fix $\gamma_{max}$ = 10$^{8}$ throughout the paper.  In Fig.~1(a), we plot $P_{1,K}(s)$ defined by Eq.~(\ref{eq2c-2}) as a function of $s$ for a typical value $\sigma=2.5$.  The solid curve, dash-dotted curve, dashed curve, and dotted curve correspond to $\gamma_{min}$ = 10, 10$^{2}$, 10$^{3}$, and 10$^{4}$, respectively.  It can be seen that the height of $P_{1,K}(s)$ is independent of $\gamma_{min}$.  In Fig.~1(b), we plot the same curves as a function of new variable $s_{C}$ which is defined by
\begin{eqnarray}
&&\hspace{-10mm}
s_{C}= s - 2 \ln 2 \gamma_{min}
\label{eq2c-18}  \, .
\end{eqnarray}
In Fig.~1(b) the four curves are totally indistinguishable, which exhibits a scaling law for $P_{1,K}(s)$.  The reason for this scaling law is as below.  For large $\gamma_{min} \gg 1$, as shown by Figs.~1(a), 1(b), and Eqs.~(\ref{eq2c-16}) and (\ref{eq2c-17}), the probability distribution function $P_{1,K}(s)$ is dominated by $P_{K,C}(s_{C},0)$, i.e. by the Compton scattering process.

\begin{figure}
\begin{center}
\includegraphics[angle=0,width=0.48\textwidth]{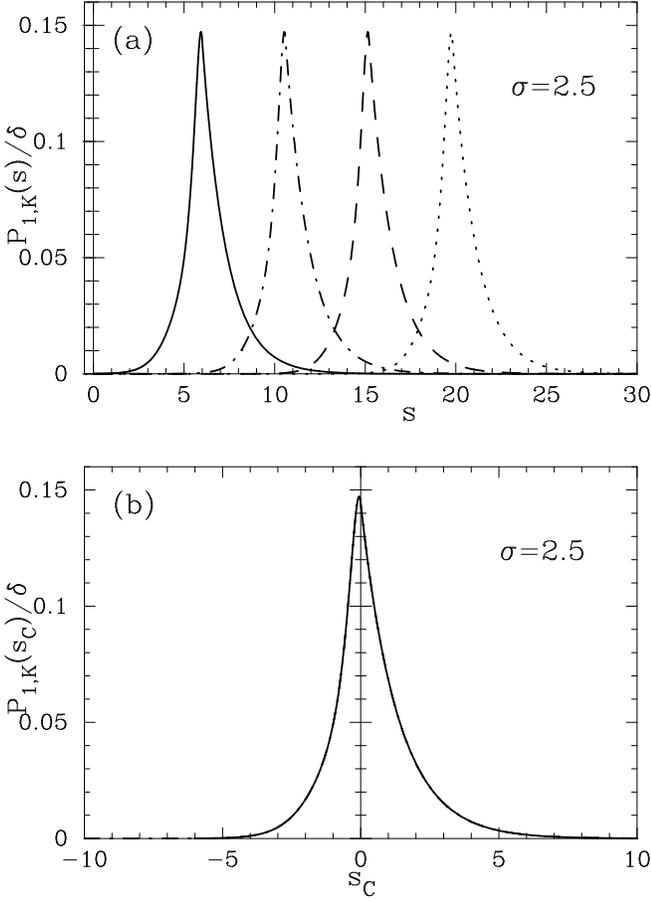}
\end{center}
\caption{Plotting of $P_{1,K}(s)$ and $P_{1,K}(s_{C})$ for $\sigma=2.5$.  Figures 1(a) and 1(b) are $P_{1,K}(s)$ and $P_{1,K}(s_{C})$, respectively.  The solid curve, dash-dotted curve, dashed curve, and dotted curve correspond to $\gamma_{min}$ = 10, 10$^{2}$, 10$^{3}$, and 10$^{4}$, respectively.}
\end{figure}

  Before closing this subsection, we study the $\sigma$ dependences on the peak position $s_{peak}$ and peak height $P_{1,K}(s_{peak})$.  As shown in Figs.\ 1(a) and 1(b), the $\gamma_{min}$ dependence of $P_{1,K}(s)$ is described by Eq.\ (\ref{eq2c-18}), namely, $s$ = $s_{C} + 2 \ln 2 \gamma_{min}$.  Therefore, we define the peak position by
\begin{eqnarray}
&&\hspace{-10mm}
s_{peak} =  s_{K}(\sigma) + 2 \ln 2\gamma_{min}
\label{eq2c-19} \, ,
\end{eqnarray}
where $s_{K}(\sigma)$ depends only on $\sigma$.  The peak position is calculated by solving the equation
\begin{eqnarray}
&&\hspace{-10mm}
\left.
\frac{ \partial  P_{1,K}(s)}{ \partial s} \right|_{s_{peak}} = 0
\label{eq2c-20}  \, .
\end{eqnarray}
  The analytic expressions for $s_{K}(\sigma)$ in the first-order and third-order approximations are given as follows:
\begin{eqnarray}
&&\hspace{-10mm}
s_{K,1st}(\sigma) = -\frac{(\sigma-1)(\sigma^{2}+ 4\sigma + 11)}{(\sigma-3)(5 \sigma^{2} + 24\sigma + 43)}
\label{eq2c-21}  \, , \\
&&\hspace{-10mm}
s_{K,3rd}(\sigma) = -\frac{1}{25\sigma + 101} \Biggl[ 13\sigma + 41
\nonumber  \\
&&\hspace{8mm}
+ \left( \frac{ \sqrt{(\sigma+3) A^{3} + B^{2}} - B }{ (\sigma+3)^{2}} \right)^{1/3} \nonumber  \\
&&\hspace{8mm}
\left. - \left( \frac{ \sqrt{(\sigma+3) A^{3} + B^{2}} + B }{ (\sigma+3)^{2}} \right)^{1/3} \right]
\label{eq2c-22}  \, ,
\end{eqnarray}
\begin{eqnarray}
&&\hspace{-5mm}
A = 81 \sigma^{3} + 637 \sigma^{2} + 2119 \sigma + 3643
\label{eq2c-23}  \, , \\
&&\hspace{-5mm}
B = 803 \sigma^{5} + 18351 \sigma^{4} + 148554 \sigma^{3} + 590290 \sigma^{2}
\nonumber \\
&&\hspace{12mm}
 + 1282323 \sigma + 1318911
\label{eq2c-24} \, .
\end{eqnarray}
We also solved Eq.~(\ref{eq2c-20}) numerically and obtained the numerical solution $s_{K,num}(\sigma)$.  In Figs.~2(a) and 2(b), we plot $s_{K}(\sigma)$ and $P_{1,K}(s_{peak})$, respectively.  The dashed curve, dash-dotted curve, and solid curve correspond to $s_{K,1st}(\sigma)$, $s_{K,3rd}(\sigma)$ and $s_{K,num}(\sigma)$, respectively.  In Fig.~2(a), the solid curve, and dash-dotted curve are indistinguishable, and three curves are almost indistinguishable in Fig.~2(b).  It can be seen from Figs.~2(a) and 2(b) that the third-order approximation is sufficiently accurate for the present purposes.

\begin{figure}
\begin{center}
\includegraphics[angle=0,width=0.48\textwidth]{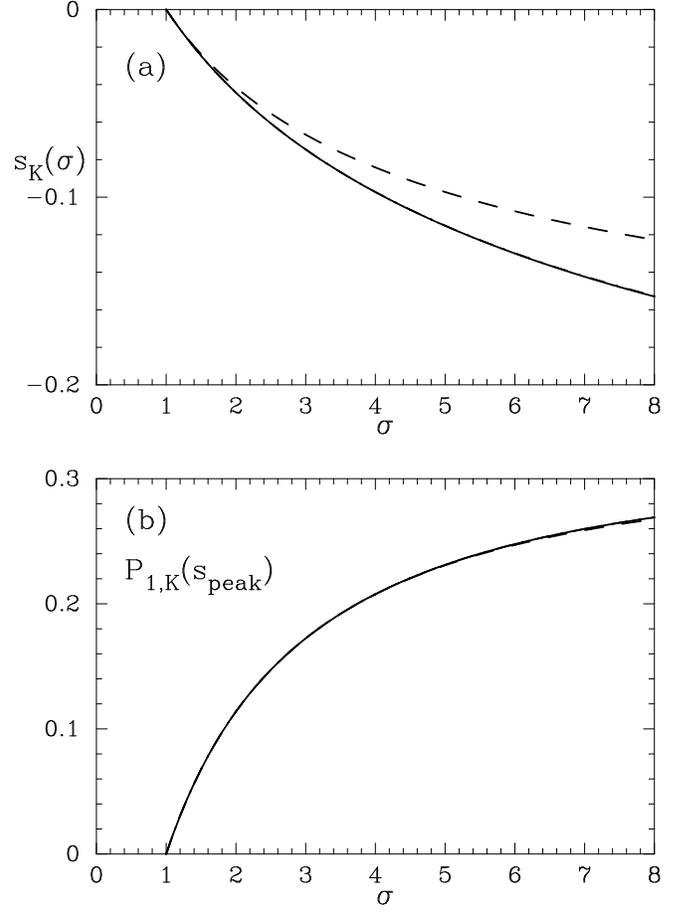}
\end{center}
\caption{Plotting of $s_{K}(\sigma)$ and $P_{1,K}(s_{peak})$.  Figures 2(a) and 2(b) are $s_{K}(\sigma)$ and $P_{1,K}(s_{peak})$, respectively.  The dashed curve, dash-dotted curve, and solid curve correspond to the first-order approximation, third-order approximation, and numerical solution, respectively.}
\end{figure}

\subsection{Scaling law for spectral intensity function}

  Let us now solve the rate equations of Eqs.~(\ref{eq2a-3}) and (\ref{eq2a-4}) with the results for $P_{1}(s)$ and $P_{1,K}(s)$.  First, one can rewrite Eq.~(\ref{eq2a-4}) as follows:
\begin{eqnarray}
&&\hspace{-10mm}
\frac{\partial I(x)}{\partial \tau} = \frac{\partial I_{I}(x)}{\partial \tau} + \beta_{C,z} \frac{\partial I_{K}(x)}{\partial \tau}
\nonumber \, , \\
&&\hspace{1mm}
 = \left( 1 - \beta_{C,z} \delta \right) \frac{\partial I_{I}(x)}{\partial \tau} + \beta_{C,z} \delta \, \frac{\partial \tilde{I}_{K}(x)}{\partial \tau}
\label{eq2d-1}  \, ,
\end{eqnarray}
where $\partial I_{I}(x)/\partial \tau$ is the result obtained in paper I, and $\partial I_{K}(x)/\partial \tau$ corresponds to non-zero bulk motions.  In deriving Eq.~(\ref{eq2d-1}), we used the following relation:
\begin{eqnarray}
&&\hspace{-10mm}
\frac{\partial I_{K}(x)}{\partial \tau} = \delta \left[ \frac{\partial \tilde{I}_{K}(X)}{\partial \tau} - \frac{\partial I_{I}(X)}{\partial \tau} \right]
\label{eq2d-2}  \, .
\end{eqnarray}
Therefore, our main concern in this section is to calculate $\partial \tilde{I}_{K}(x)/\partial \tau$, which is defined as follows:
\begin{eqnarray}
&&\hspace{-10mm}
\frac{\partial \tilde{I}_{K}(x)}{\partial \tau}
 = \int_{-\infty}^{\infty}ds \tilde{P}_{1,K}(s)
\left[e^{-s} I_{0}(e^{-s}x)- I_{0}(x)\right] \, ,
\label{eq2d-3}
\end{eqnarray}
where $I_{0}(x)=I_{0}x^{3}/(e^{x}-1)$ and $I_{0}=(k_{B}T_{CMB})^{3}/2\pi^{2}$, because we consider the CMB photons for the initial distribution.  It should be noted that the useful relation $\tilde{P}_{1,K}(s)e^{-4s}=\tilde{P}_{1,K}(-s)$ was used in deriving Eq.~(\ref{eq2d-3}).

  Similarly, Eq.~(\ref{eq2a-3}) can be rewritten as follows:
\begin{eqnarray}
&&\hspace{-10mm}
\frac{\partial n(x)}{\partial \tau} = \left( 1 - \beta_{C,z} \delta \right) \frac{\partial n_{I}(x)}{\partial \tau} + \beta_{C,z} \delta \, \frac{\partial \tilde{n}_{K}(x)}{\partial \tau}
\label{eq2d-4}  \, ,
\end{eqnarray}
where $\partial n_{I}(x)/\partial \tau$ is the result obtained in paper I, and $\partial \tilde{n}_{K}(x)/\partial \tau$ is defined as follows:
\begin{eqnarray}
&&\hspace{-10mm}
\frac{\partial \tilde{n}_{K}(x)}{\partial \tau}
 = \int_{-\infty}^{\infty}ds \tilde{P}_{1,K}(s)
\left[n_{0}(e^sx)- n_{0}(x)\right] \, ,
\label{eq2d-5}  \\
&&\hspace{+3mm}
= \frac{1}{I_{0} x^{3}} \frac{\partial \tilde{I}_{K}(x)}{\partial \tau}
\label{eq2d-6}  \, ,
\end{eqnarray}
where $n_{0}(x)=1/(e^{x}-1)$.

  For the inverse Compton scattering of the CMB photons off high-energy electrons, we are interested in high-energy spectra such as X-rays ($\sim$ keV) and gamma-rays ($\sim$ MeV).  Therefore, one can safely assume
\begin{eqnarray}
&&\hspace{-10mm}
   x \equiv \frac{\omega}{k_{B}T_{CMB}} \gg 1
\label{eq2d-7}
\end{eqnarray}
for scattered photons.  For the $\gamma$ parameters, we assume the same condition used in the scaling law for $P_{1,K}(s)$, namely,
\begin{eqnarray}
&&\hspace{-10mm}
1 \ll \gamma_{min} \ll \gamma_{max}  \, .
\label{eq2d-8}
\end{eqnarray}
Under these assumptions, Eqs.~(\ref{eq2d-3}) and (\ref{eq2d-5}) are much simplified, and can be solved analytically.  In Appendix B of paper I, we have shown the derivation for $\partial I_{I}(x)/\partial \tau$ and $\partial n_{I}(x)/\partial \tau$ in detail.  One can also calculate $\partial \tilde{I}_{K}(x)/\partial \tau$ and $\partial \tilde{n}_{K}(x)/\partial \tau$ in a similar manner.

  The final results are as follows:
\begin{eqnarray}
&&\hspace{-10mm}
\frac{d \tilde{I}_{K}(X)}{d \tau}= \frac{1}{x} 3(\sigma-1) I_0
\left[ X^3\int_{X}^{\infty} dt \frac{1}{e^t-1}
\right.
\nonumber \\
&&\hspace{-13mm}
\times
\left\{ \frac{1}{\sigma+5} \left( \frac{3\sigma+23}{\sigma+5}+4\ln\frac{t}{X} \right) - \frac{4}{\sigma+3} \frac{t}{X} + \frac{1}{\sigma+1}\frac{t^{2}}{X^{2}} \right\}
\nonumber \\
&&\hspace{-13mm}
\left. + \, \frac{4(\sigma^2+4\sigma+11)}{(\sigma+1)(\sigma+3)(\sigma+5)^{2}} \frac{1}{X^{(\sigma-1)/2}}
 \int_{0}^{X}dt\frac{t^{(\sigma+5)/2}}{e^t-1}
\right]
\label{eq2d-9}  \, ,
\end{eqnarray}
\begin{eqnarray}
&&\hspace{-20mm}
\frac{d \tilde{n}_{K}(X)}{d \tau} = \frac{1}{64 \gamma_{min}^{6}} \frac{1}{I_{0}X^{3}} \frac{d \tilde{I}_{K}(X)}{d \tau}  \, ,
\label{eq2d-10}
\end{eqnarray}
where $X$ is the scaling variable defined by
\begin{eqnarray}
&&\hspace{-10mm}
X = \frac{x}{4 \gamma_{min}^{2}}
\label{eq2d-11}  \, .
\end{eqnarray}
In order to compare the present results for $d \tilde{I}_{K}(X)/d \tau$ and $d \tilde{n}_{K}(X)/d \tau$ with $d I_{I}(X)/d \tau$ and $d n_{I}(X)/d \tau$, we recall the results of paper I.  They are given as follows:
\begin{eqnarray}
&&\hspace{-8mm}
\frac{d I_{I}(X)}{d \tau}= 3(\sigma-1) I_0
\left[ X^3\int_{X}^{\infty} \frac{dt}{t} \frac{1}{e^t-1}
\right.
\nonumber \\
&&\hspace{-8mm}
\times
\left\{ -\frac{2}{\sigma+5}
+\frac{1}{\sigma+3}\left(\frac{\sigma-1}{\sigma+3}-2\ln\frac{t}{X}\right) \frac{t}{X} +\frac{1}{\sigma+1}\frac{t^{2}}{X^{2}} \right\}
\nonumber \\
&&\hspace{-8mm}
\left. + \, \frac{2(\sigma^2+4\sigma+11)}{(\sigma+1)(\sigma+3)^2(\sigma+5)}
\frac{1}{X^{(\sigma-1)/2}}
 \int_{0}^{X}dt\frac{t^{(\sigma+3)/2}}{e^t-1}
\right]
\label{eq2d-12}  \, ,
\end{eqnarray}
\begin{eqnarray}
&&\hspace{-25mm}
\frac{d n_{I}(X)}{d \tau} = \frac{1}{64 \gamma_{min}^{6}} \frac{1}{I_{0}X^{3}} \frac{d I_{I}(X)}{d \tau}  \, .
\label{eq2d-13}
\end{eqnarray}

  Comparing Eqs.~(\ref{eq2d-9}) and (\ref{eq2d-10}) with Eqs.~(\ref{eq2d-12}) and (\ref{eq2d-13}), respectively, one finds as follows:
\begin{eqnarray}
&&\hspace{-10mm}
\frac{d \tilde{I}_{K}(X)}{d \tau} \approx \frac{1}{x} \frac{d I_{I}(X)}{d \tau}
\label{eq2d-14}  \, ,  \\
&&\hspace{-10mm}
\frac{d \tilde{n}_{K}(X)}{d \tau} \approx \frac{1}{x} \frac{d n_{I}(X)}{d \tau}
\label{eq2d-15} \, .
\end{eqnarray}
Thus, it is found that $d\tilde{I}_{K}(X)/d\tau$ and $d\tilde{n}_{K}/d\tau$ are suppressed by a factor $1/x$ compared with $dI_{I}(X)/d\tau$ and $dn_{I}/d\tau$, respectively, and they are negligible for $x \gg 1$.  Finally, one obtains the total contribution as follows:
\begin{eqnarray}
&&\hspace{-10mm}
\frac{d I(X)}{d \tau} = \left(1 - \beta_{C,z} \delta \right) \frac{d I_{I}(X)}{d \tau}
\label{eq2d-16}  \, ,  \\
&&\hspace{-10mm}
\frac{d n(X)}{d \tau} = \left(1 - \beta_{C,z} \delta \right) \frac{d n_{I}(X)}{d \tau}
\label{eq2d-17} \, .
\end{eqnarray}

  As seen from Eqs.~(\ref{eq2d-16}) and (\ref{eq2d-17}), the same scaling-law shown in paper I is valid to the present case which includes the effect of the bulk motions.  Moreover, it should be noted that the effect of the bulk motions is approximately expressed by the factor $-\beta_{C,z} \delta$, where $\beta_{C,z}$ is the bulk velocity parallel to the observer, $\delta=\sigma+2$, and $\sigma$ is the power-index of the power-law electron distribution function.  It turns out that the effect of the bulk motions is small for most of the astrophysical objects.  For example, $\beta_{C,z} \approx 1/300$ for a typical CG and $\delta=6$ for a typical value of $\sigma=4$, one has $-\beta_{C,z} \delta \approx - 2\%$.  Therefore, the astrophysical applications of the scaling laws suggested in paper I are still fully valid even for the astrophysical objects which have the bulk motions.

\section{Effect of observer's motion}

\subsection{Kinematics}

  In the present section, let us discuss the effect of the observer's motion.  
It is well known that the Solar System has a bulk motion with respect to the CMB frame.  Assuming that the CMB dipole is fully motion-induced, we deduce that the Solar System is moving with a velocity $\beta_{S} \equiv v_{S}/c$ = 1.241$\times10^{-3}$ towards the direction $(\ell, b)=(264.14^{\circ} \pm 0.15^{\circ}$, $48.26^{\circ} \pm 0.15^{\circ}$)\cite{smoo77, fixs96, fixs02}.  Chluba et al.\cite{chlu05} calculated corrections to the SZ effect for the CG arising from the bulk motion of the Solar System.  Nozawa, Itoh and Kohyama\cite{noza05} calculated the effect in more general way with the Lorentz covariant formalism.  In the present paper, we apply the method developed by Nozawa, Itoh and Kohyama\cite{noza05} to the high-energy inverse Compton scattering and calculate the effect of the bulk motion of the observer.(Readers may be referred to the paper for the details.)

  Let us suppose that the observer's system (the Solar System) is moving with a velocity $\vec{\beta}_{S}(\equiv \vec{v}_{S}/c)$ with respect to the CMB.  The $z$ axis is fixed to a line connecting the observer and the center of mass of the CG as defined in Sec.~II.  Note that variables in the Solar System will be denoted by the subscript $S$, unless otherwise stated explicitly.  First, the photon energies $\omega$ and $\omega_{S}$ are related by the Lorentz transformation:
\begin{eqnarray}
&&\hspace{-10mm}
\omega = \gamma_{S} \omega_{S}(1 - \beta_{S} \mu_{S})
\label{eq3a-1} \, ,  \\
&&\hspace{-12mm}
\gamma_{S} = \frac{1}{ \sqrt{1 - \beta_{S}^{2} }} 
\label{eq3a-2}  \, ,
\end{eqnarray}
where $\mu_{S}= - \hat{\beta}_{S} \cdot \hat{k}_{S}$, and $\hat{\beta}_{S}$ is a unit vector in the direction of $\vec{\beta}_{S}$.  The photon unit wave vectors $\hat{k}$ and $\hat{k}_{S}$ are related by\cite{mell62}
\begin{eqnarray}
&&\hspace{-10mm}
\hat{k}  =  \left(\frac{ \hat{k}_{S} \cdot \hat{\beta}_{S} + \beta_{S} }{1 + \hat{k}_{S} \cdot \vec{\beta}_{S} } \right) \hat{\beta}_{S} +   \frac{ \hat{k}_{S} - ( \hat{k}_{S} \cdot \hat{\beta}_{S} ) \hat{\beta}_{S} }{ \gamma_{S}(1 + \hat{k}_{S} \cdot \vec{\beta}_{S} ) } 
\label{eq3a-3}  \, .
\end{eqnarray}
Similarly, the velocity of the CG in the CMB frame $\vec{\beta}_{C}$ is related by the velocity of the CG in the Solar System $\vec{\beta}_{C}^{\prime}$ by the following relationship\cite{mell62}:
\begin{eqnarray}
&&\hspace{-10mm}
\vec{\beta}_{C}  =  \left(\frac{ \vec{\beta}_{C}^{\prime} \cdot \hat{\beta}_{S} + \beta_{S} }{1 + \vec{\beta}_{C}^{\prime} \cdot \vec{\beta}_{S} } \right) \hat{\beta}_{S} +   \frac{ \vec{\beta}_{C}^{\prime} - ( \vec{\beta}_{C}^{\prime} \cdot \hat{\beta}_{S} ) \hat{\beta}_{S} }{ \gamma_{S}(1 + \vec{\beta}_{C}^{\prime} \cdot \vec{\beta}_{S} ) } 
\label{eq3a-4}  \, .
\end{eqnarray}
Thus, the kinematics for the CMB, CG and Solar System are well defined.  Then, the initial (Planckian) photon distribution functions in the CMB frame and the Solar System are related as follows\cite{noza05}:
\begin{eqnarray}
&&\hspace{-18mm}
n(\omega, \vec{k}) = n_{S}(\omega_{S}, \vec{k}_{S})
\label{eq3a-5} \, .
\end{eqnarray}

  For the practical calculation we now assume a condition $\beta_{S} \ll 1$, which is well satisfied.  Therefore we keep only the first-order terms in $\beta_{S}$ in the calculations throughout the present paper.  In this approximation, Eqs.~(\ref{eq3a-1}), (\ref{eq3a-3}) and (\ref{eq3a-4}) are much simplified as follows:
\begin{eqnarray}
&&\hspace{-10mm}
\omega = \omega_{S}(1 - \beta_{S} \mu_{S})
\label{eq3a-6} \, ,  \\
&&\hspace{-10mm}
\hat{k} = \hat{k}_{S}
\label{eq3a-7}  \, , \\
&&\hspace{-10mm}
\vec{\beta}_{C} = \vec{\beta}_{C}^{\prime} + \vec{\beta}_{S}
\label{eq3a-8}  \, .
\end{eqnarray}
Inserting Eqs.~(\ref{eq3a-6}) and (\ref{eq3a-7}) into Eq.~(\ref{eq3a-5}), one obtains as follows:
\begin{eqnarray}
&&\hspace{-10mm}
n(x) = n_{S}(x_{S})
\nonumber  \\
&&\hspace{-2mm}
 = \frac{1}{e^{x_{S}(1- \beta_{S} \mu_{S})}-1}
\nonumber  \\
&&\hspace{-2mm}
= \frac{1}{e^{x_{S}}-1} - \beta_{S,z} \frac{x_{S}e^{x_{S}}}{\left(e^{x_{S}}-1\right)^{2}} + O(\beta_{S}^{2})
\label{eq3a-9}  \, ,
\end{eqnarray}
where $x_{S}=\omega_{S}/k_{B}T_{CMB}$, and $\beta_{S} \mu_{S}= -\beta_{S,z}$.  In Eq.~(\ref{eq3a-9}), the second term is the first-order correction due to the bulk motion of the observer.

\subsection{Scaling law for spectral intensity function}

  Now, our task in the present section is to derive the analytic expressions for the spectral intensity function and photon number distribution function with variables in the observer's system.  This can be done by inserting Eq.~(\ref{eq3a-9}) into Eqs.~(\ref{eq2d-16}) and (\ref{eq2d-17}).  One finally obtains the following expressions which include the effect of the bulk motions for both the astrophysical object (CG) and observer (Solar System):
\begin{eqnarray}
&&\hspace{-10mm}
\frac{d I(X_{S})}{d \tau} = \left(1 - \beta_{C,z} \delta \right) \frac{d I_{I}(X_{S})}{d \tau} - \beta_{S,z} \frac{d I_{S}(X_{S})}{d \tau}
\label{eq3b-1}  \, ,  \\
&&\hspace{-10mm}
\frac{d n(X_{S})}{d \tau} = \left(1 - \beta_{C,z} \delta \right) \frac{d n_{I}(X_{S})}{d \tau} - \beta_{S,z} \frac{d n_{S}(X_{S})}{d \tau}
\label{eq3b-2} \, ,
\end{eqnarray}
where $X_{S}=x_{S}/4\gamma_{min}^{2}$.  Note that $\beta_{C,z}$ is the bulk velocity of the CG in the CMB frame.  The explicit forms for $dI_{S}(X_{S})/d\tau$ and $dn_{S}(X_{S})/d\tau$ are given as follows:
\begin{eqnarray}
&&\hspace{-9mm}
\frac{d I_{S}(X_{S})}{d \tau}= 3(\sigma-1) I_0
\left[ X_{S}^3\int_{X_{S}}^{\infty} dt\frac{e^{t}}{(e^t-1)^{2}}
\right.
\nonumber \\
&&\hspace{-9mm}
\times
\left\{ -\frac{2}{\sigma+5}
+\frac{1}{\sigma+3}\left(\frac{\sigma-1}{\sigma+3}-2\ln\frac{t}{X_{S}}\right) \frac{t}{X_{S}} +\frac{1}{\sigma+1}\frac{t^{2}}{X_{S}^{2}} \right\}
\nonumber \\
&&\hspace{-9mm}
\left. + \, \frac{2(\sigma^2+4\sigma+11)}{(\sigma+1)(\sigma+3)^2(\sigma+5)}
\frac{1}{X_{S}^{(\sigma-1)/2}}
 \int_{0}^{X_{S}}dt\frac{t^{(\sigma+5)/2}e^{t}}{(e^t-1)^{2}}
\right]
\label{eq3b-3}  \, ,
\end{eqnarray}
\begin{eqnarray}
&&\hspace{-25mm}
\frac{d n_{S}(X_{S})}{d \tau} = \frac{1}{64 \gamma_{min}^{6}} \frac{1}{I_{0}X_{S}^{3}} \frac{d I_{S}(X_{S})}{d \tau}  \, .
\label{eq3b-4}
\end{eqnarray}

  As seen from Eqs.~(\ref{eq3b-1}) and (\ref{eq3b-3}), the spectral intensity function has the scaling law as the previous case even after including the effect of the bulk motions for both the astrophysical object (CG) and observer (Solar System).  Therefore, it should be emphasized that the scaling law is universal.

  Furthermore, the effect of the bulk motion of the observer is small compared with the leading-order term because of the smallness of $\beta_{S}$.  In Fig.~3, we plot $dI_{I}(X_{S})/d\tau$ and $\beta_{S,z}dI_{S}(X_{S})/d\tau$ for $\beta_{S,z}=1.241\times10^{-3}$ as a function of $X_{S}$.  In the calculation, a typical value $\sigma=3.5$ was used for an illustrative purpose.  It is clear from Fig.~3 that the effect is quite small.  The correction is roughly 0.4\% at the peak position.

\begin{figure}
\begin{center}
\includegraphics[angle=0,width=0.48\textwidth]{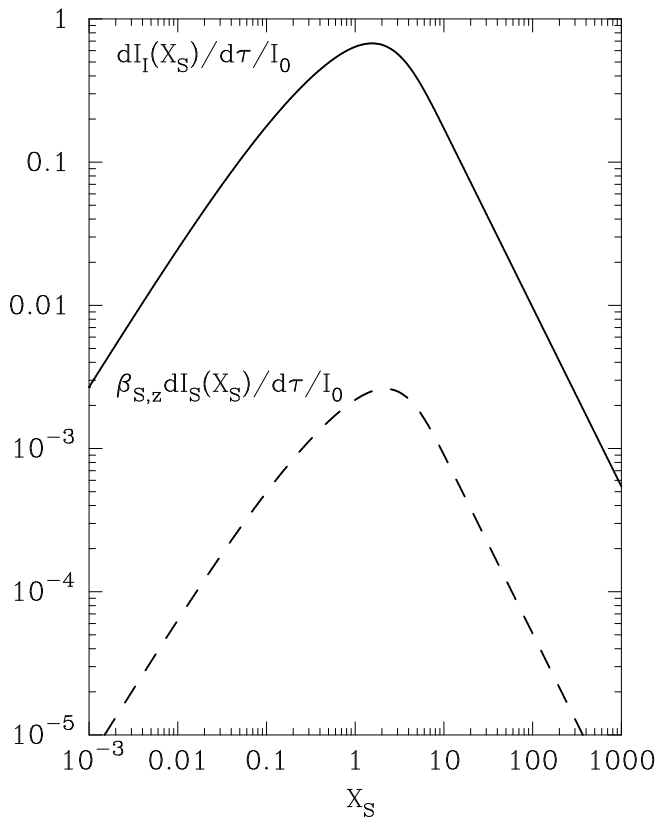}
\end{center}
\caption{Plotting of $dI_{I}(X_{S})/d\tau$ and $\beta_{S,z}dI_{S}(X_{S})/d\tau$ for $\beta_{S,z}=1.241\times10^{-3}$ as a function of $X_{S}$.  The solid curve is $dI_{I}(X_{S})/d\tau$ and the dashed curve corresponds to $\beta_{S,z}dI_{S}(X_{S})/d\tau$.  In the calculation, a typical value $\sigma=3.5$ was used for an illustrative purpose.}
\end{figure}

\section{Concluding Remarks}

  In paper I\cite{noza10}, we studied the high-energy inverse Compton scattering of the CMB photons off nonthermal electrons with the formalism derived in the Thomson approximation\cite{noza09a}.  As for the nonthermal electron distribution function, a standard power-law distribution function of three parameters were adopted: the power-index $\sigma$, minimum value $\gamma_{min}$, and maximum value $\gamma_{max}$ of the distribution range.  For the case $\gamma_{min} \gg 1$, a scaling law in the probability distribution function $P_{1}(s)$ were found, where the peak position depends on $s- 2\ln 2\gamma_{min}$, and the peak height depends only on the power-index parameter.  The spectral intensity function $I_{I}(x)$ was also calculated.  For the case of high-energy photons of $x \gg 1$, a scaling law in $dI_{I}(x)/d\tau$ were found, where the function depends on a new variable $X=x/(4\gamma_{min}^{2})$.  The peak position and peak height depend only on the power-index parameter.  The $\gamma_{min}$ dependence of $dI_{I}(X)/d\tau$ is included in the variable $X$.  The formalism was applied to the observations of the spectral intensity function in the X-ray and gamma-ray energy regions.  It was found that the observations in the X-ray and gamma-ray regions had sensitivities of $\gamma_{min}$=500 $\sim$ 3$\times 10^{3}$ and $\gamma_{min}$=16$\times 10^{3}$ $\sim$ 95$\times 10^{3}$, respectively.

  In the present paper, we have extended the formalism obtained in paper I to the cases where the astrophysical objects have the bulk motions with respect to the CMB frame (, for example, the peculiar velocity of the CG).  The extension will be particularly interesting for the analysis of X-ray and gamma-ray emissions, for example, from radio galaxies with non-zero peculiar velocities and various astrophysical jets.

  First, we have derived the analytic expressions for the redistribution functions $P_{K}(s,\gamma)$ and $P_{1,K}(s)$.  Assuming the power-law electron distribution, we have shown that $P_{1,K}(s)$ also has the same scaling law as $P_{1}(s)$, where the peak height and peak position depend only on the power-index parameter.

  Then, we have calculated the rate equations and obtained the analytic expressions for $dI(X)/d\tau$ and $dn(X)/d\tau$ which include the effect of the bulk motions.  It has been found that the same scaling-law shown in paper I is valid to the present case which includes the effect of the bulk motions, where the peak height and peak position depend only on the power-index parameter.

  It has been found that the effect of the bulk motions is approximately expressed by the factor $-\beta_{C,z} \delta$, where $\beta_{C,z}$ is the bulk velocity parallel to the observer and $\delta=\sigma+2$ is a factor related to the electron distribution function.  It has been shown that the effect of the bulk motions is small for most of the astrophysical objects.  For example, $\beta_{C,z} \approx 1/300$ for a typical CG and $\delta=6$ for a typical value of $\sigma=4$, one has $-\beta_{C,z} \delta \approx - 2\%$.

  We have also calculated the effect of the bulk motion of the Solar System by applying the method developed by Nozawa, Itoh, and Kohyama\cite{noza05}.  The analytic expression for the spectral intensity function and photon number distribution function have been derived.  It has been found that the same scaling laws is valid for the spectral intensity function.  Furthermore, it has been found that the effect of the observer's motion is small compared with the leading-order term because of the smallness of the velocity $\beta_{S}$.  The correction is roughly 0.4\% at the peak position for a typical value of $\sigma=3.5$.

  In conclusion, the astrophysical applications of the scaling laws suggested in paper I are still fully valid even if the astrophysical object and observer have the bulk motions.

\begin{acknowledgments}
This work is financially supported in part by the Grant-in-Aid of Japanese Ministry of Education, Culture, Sports, Science, and Technology under Contract No. 21540277.  We would like to thank our referee for valuable suggestions.
\end{acknowledgments}


\bibliography{apssamp}

\end{document}